\documentclass{article}
\usepackage{makecell}

\usepackage{amsmath}

\usepackage{PRIMEarxiv}

\usepackage[utf8]{inputenc} 
\usepackage[T1]{fontenc}    
\usepackage{hyperref}       
\usepackage{url}            
\usepackage{booktabs}       
\usepackage{amsfonts}       
\usepackage{nicefrac}       
\usepackage{microtype}      
\usepackage{lipsum}
\usepackage{fancyhdr}       
\usepackage{graphicx}       
\graphicspath{{media/}}     

\title{The Gender Pay Gap in China: Insights from a Discrimination Perspective
}

\author{
    Wei Bai$^{a,1}$, Yan-Li Lee$^b$, Jingyi Liao$^c$, Lusi Wu$^{d,2}$, Mei Xie$^{a,1}$, Tao Zhou$^{a,1,2}$\thanks{\textsuperscript{1}These three authors contribute equally to this work. \newline \indent  \textsuperscript{\hspace{0.8em}} \textsuperscript{2}To whom correspondence should be addressed. E-mail: wulusi@uestc.edu.cn (L. W.); zhutou@ustc.edu (T. Z.)}\\
  $^a$ CompleX Lab, University of Electronic Science and Technology of China, Chengdu 611731\\
  $^b$ School of Computer and Software Engineering, Xihua University, Chengdu 610039\\
  $^c$ Shenzhen International Graduate School, Tsinghua University, Shenzhen 518055\\
  $^d$ School of Management and Economics, University of Electronic Science and Technology of China, Chengdu 611731\\
}

\makeatletter
\def\thanks#1{\protected@xdef\@thanks{\@thanks
        \protect\footnotetext{#1}}}
\makeatother

\begin{document}
\maketitle

\begin{abstract}
Equal pay is an essential component of gender equality, one of the Sustainable Development Goals of the United Nations. Using resume data of over ten million Chinese online job seekers in 2015, we study the current gender pay gap in China. The results show that on average women only earned 71.57\% of what men earned in China. The gender pay gap exists across all age groups and educational levels. Contrary to the commonly held view that developments in education, economy, and a more open culture would reduce the gender pay gap, the fusion analysis of resume data and socio-economic data presents that they have not helped reach the gender pay equality in China. China seems to be stuck in a place where traditional methods cannot make further progress. Our analysis further shows that 81.47\% of the variance in the gender pay gap can be potentially attributed to discrimination. In particular, compared with the unmarried, both the gender pay gap itself and proportion potentially attributed to discrimination of the married are larger, indicating that married women suffer greater inequality and more discrimination than unmarried ones. Taken together, we suggest that more research attention should be paid to the effect of discrimination in understanding gender pay gap based on the family constraint theory. We also suggest the Chinese government to increase investment in family-supportive policies and grants in addition to female education.
\end{abstract}


\section{Introduction}
As one of the Sustainable Development Goals of the United Nations, gender equality is not only a basic human right, but also an essential element for peace, prosperity, and sustainable development of the world \cite{world2011world}. Reducing gender inequality has a long-term positive impact on economic growth \cite{ahang2014impact}. Gender inequality is mainly evaluated along four dimensions: economic status, learning opportunities, political participation, and health welfare. Given that gender inequality in pay constitutes the most prominent part of economic status, reducing gender pay gap is of vital importance in achieving gender equality. However, the country-level gender pay gap still persists around the world. Using the ratio of women's earnings to men's as a quantitative index, the female-male ratio, on average, is about 75\% in hourly pay and 80\% in monthly pay globally \cite{international2021global}. In China, the gender pay gap is also hard to ignore. Table \ref{tab1} summarizes the main findings of research on the gender pay gap in China using data collected from a urban household questionnaire. As shown in the table, the level of gender pay gap in China is smaller than the global average around the 1990s, but has trended upward (corresponding to a decreasing female-male ratio) in the last two decades. The current gender pay gap in China is already higher than the global average.

\begin{table}[tbhp]
\centering
\small
\caption{The female-male ratios of salary in China, calculated from the China Family Panel Studies (CFPS) in different years.}
\begin{tabular}{cccccc}
    \toprule
    References & 1988 & 1995 & 2002 & 2007 & 2013 \\
    \midrule
    \cite{bishop2005economic} & 81\% & 80\% & - & - & - \\
    \cite{demurger2007evolution} & 81.3\% & 82.2\% & - & - & - \\
    \cite{gustafsson2000economic} & 84.4\% & 82.5\% & - & - & - \\
    \cite{shi2011evolution} & - & 84\% & 82\% & 74\% & - \\
    \cite{song2017china} & - & 87\% & 82\% & 71\% & 75\% \\
    \bottomrule
\end{tabular}
\label{tab1}
\end{table}

Economists have proposed two main explanations for the gender pay gap: the differences in human capital and the discrimination in the labor market (i.e., the difference in wages between men and women who possess equal human capital) \cite{blau2007the}. Yet scholars and practitioners so far have primarily focused on human captial. For example, Blau \textit{et al.} \cite{blau1997swimming} claimed that the improvement of female human capital in the 1980s in the United States could explain the reduction of the gender pay gap. Gustafsson \textit{et al.} \cite{gustafsson2000economic} considered the reduced gender pay gap in China in 1995 relative to 1988 as a result of women's increased education and work experience. By analyzing the trend of gender pay in urban China from 1989 to 2004, Liu \cite{liu2011economic} concluded that women's educational attainment reduced the gender pay gap by about 4.3\%. Similarly, Wang \cite{wei2013class} found that improving women's vocational skills exerts a positive impact on gender pay equality. It is thus not surprising that the Equal Pay International Coalition (EPIC) has paid great attention to increase female human capital and pay transparency in order to achieve equal pay for equal work regardless of gender \cite{bmfsfj2020the} (see also recent analyses on the the influence of pay transparency on gender inequality \cite{obloj2022the}).

The discrimination literature, by contrast, considers gender discrimination as the main reason for the gender pay gap. It holds that the discrimination plays a more important role than human capital in causing differences in earnings between men and women \cite{pacheco2017empirical,atencio2015gender,rahman2019male,khitarishvili2013evaluating}. Supporting this notion, Lee and Wie \cite{jong2017wage} analyzed the gender pay gap between China and India from 1988 to 2010, and concluded that the severe gender discrimination in China’s labor force was the main cause. Some studies found that women and men have near-identical human capital at college exit, but cultural beliefs about men as more fit for STEM professions may lead to a significant gender pay gap \cite{sterling2020confidence,cyr2021mapping}. Unfortunately, these studies could not deepen our understanding of why and how such discrimination happens (i.e., the causes of discrimination) and the mechanism linking it with the gender pay gap. Hence extant literature provides little insights on specific strategies for gender pay equality via discrimination-reducing measures, despite the well-discussed notion that discrimination predicts gender pay gap.

Recently, natural data, data collected in a way that the targets are not aware of the fact that they are recorded and analyzed, have been increasingly used in socio-economic research \cite{gao2019computational,zhou2021representative}. Resumes submitted by job seekers on the Internet can be used to analyze pay gaps among different groups and occupations \cite{yang2018height,hangartner2021monitoring}. Using data from over ten millions of resumes of Chinese online job seekers in 2015, we study the current gender pay gap in China. We not only present the gender pay gap, but also examine its sources, and discuss the challenges and possible solutions.

\section{Results}
We use resume data of 10,318,484 job seekers crawled from various online recruitment websites in 2015. The data includes demographic information such as gender, age, educational experience, work experience, and expected occupation. The most recent annual salary reported in the resume is used as the person's true salary. Table \ref{tab2} presents the basic statistics of the resume data. The detailed description and screening methods can be found in Materials and Methods.

\begin{table*}[tbhp]
\centering
\scriptsize
\caption{Basic statistics of the resume data. Academic degree is coded as 4 for Ph.D., 3 for Master, 2 for Bachelor, and 1 for Others. University from which the job seeker earned the degree is coded as 4 for 985 Universities, 3 for 211 Universities, 2 for Other Universities, and 1 for Others. Mean values of metrics are reported (ranges in brackets). CNY stands for Chinese Yuan. \newline}
\begin{tabular}{lccrrrrrr}
    \toprule
    Gender & Observations & Married/Unmarried & Age (Year) & Work Tenure (Year) & Degree & University Type & Salary (1K CNY) \\
    \midrule
    Female & 1,582,339 & 370,559/490,931 & \makecell[c]{27.3690\\$[$16, 65$]$} &  \makecell[c]{4.7078\\$[$0.0027, 53.5397$]$} & \makecell[c]{1.6647\\$[$1, 4$]$} &  \makecell[c]{2.1572\\$[$1, 4$]$} & \makecell[c]{66.5027\\$[$1, 1875$]$} \\
    Male & 2,373,962 & 655,664/671,668 & \makecell[c]{29.1395\\$[$16, 65$]$} & \makecell[c]{5.7607\\$[$0.0027, 63.6521$]$} & \makecell[c]{1.6667\\$[$1, 4$]$} & \makecell[c]{2.2067\\$[$1, 4$]$} & \makecell[c]{92.9176\\$[$1, 2000$]$} \\
    \bottomrule
\end{tabular}
\label{tab2}
\end{table*}

\begin{figure}[tbhp]
\centering
\includegraphics[width=.5\linewidth]{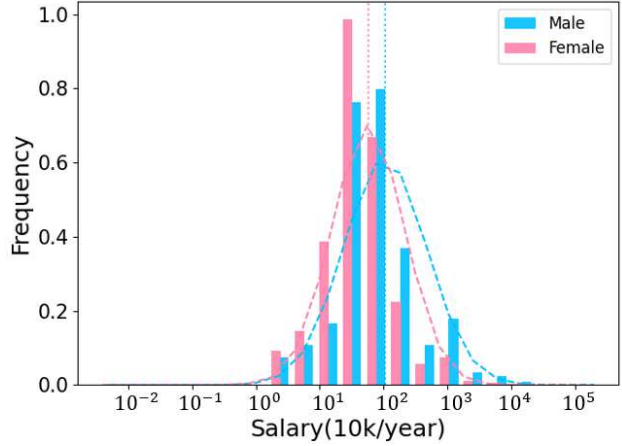}
\caption{Distributions of pay of men (blue) and women (red). The dashed curves are log-normal fits for those two distributions. The dotted vertical lines denote the average values.}
\label{fig1}
\end{figure}

We use women's salary relative to men’s (i.e., the female-male ratio) to measure the gender pay gap, a quantitative index recommended by the European Union \cite{blau2003understanding}. We find that the female-male ratio in China is 71.57\%, with an average annual pay of 92,918 CNY for men and 66,503 CNY for women. Figure \ref{fig1} shows the distributions of both men's and women's pay. It shows that the proportion of men with low pay is significantly lower than that of women. Meanwhile, the proportion of men with high pay is significantly higher than that of women. Besides, the overall pay of men is significantly higher than that of women.

\begin{figure*}
\centering
\includegraphics[width=1\linewidth]{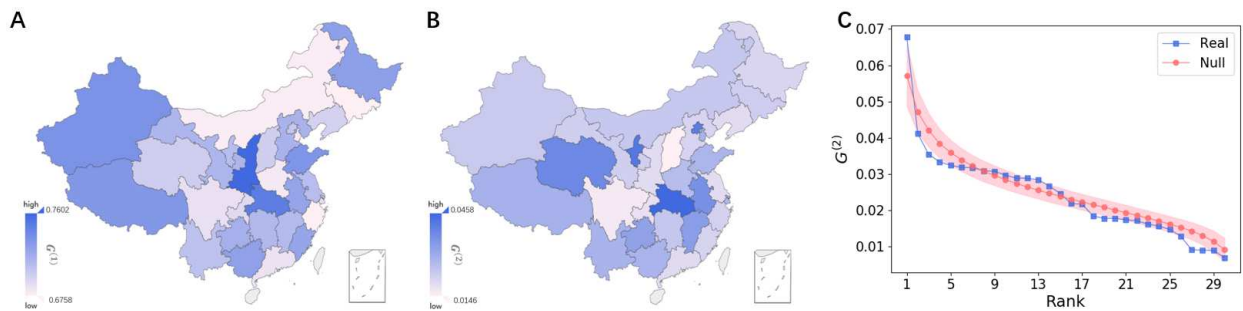}
\caption{(A) Geographical distribution of $G^{(1)}$. (B) Geographical distribution of $G^{(2)}$. (C) Comparison between real data and null model on ranked values of $G^{(2)}$. The red curve is the average over 1000 simulations and the shaded area denotes the variance.}
\label{fig2}
\end{figure*}

We also examine the gender pay gap in different provinces, and analyze the relationship between regional socio-economic development and gender pay inequality. For a province $i$, the first-order difference of gender pay gap is defined as the average pay of women relative to men based on all resume data in that province, i.e.

\begin{equation}
    G_i^{(1)} = r_i,
\label{eqn1}
\end{equation}
where $r_i$ is the female-male ratio of province $i$. Figure \ref{fig2}A shows the geographical distribution of $G^{(1)}$ at the provincial level (see Table S1 in SI Appendix for ranked $G^{(1)}$ of all provinces), where lighter colors indicate relatively larger gender pay gap. On a provincial average, women's pay in China is 70.63\% of men's ($[$67.58\%, 76.02\%$]$ with a standard deviation of 2.15\%). In general, the gender pay gap is observed in every province, and the geographical distribution of $G^{(1)}$ is relatively even.

The second-order difference of a province $i$ is defined as the mean of $G^{(1)}$ differences between $i$ and its contiguous provinces:

\begin{equation}
    G_i^{(2)} = \frac{1}{|\Gamma_i|}  \sum\limits_{j\in\Gamma_i}|G_j^{(1)}-G_i^{(1)}|,
\label{eqn2}
\end{equation}
where $\Gamma_i$ is the set of provinces contiguous to province $i$. The level of socio-economic development is clustered across regions in China. That is, the overall similarity in economic, educational, cultural, and industrial structure of neighboring provinces is higher than that of provinces farther away from each other \cite{he2008globalization,gao2021spillovers,mao2021cultural}. If the level of the gender pay gap is influenced by the level of socio-economic development, $G^{(2)}$ is expected to be very small, which means the level of the gender pay gap in neighboring provinces is similar. Figure \ref{fig2}B shows the geographical distribution of $G^{(2)}$ at the provincial level (see Table S2 in SI Appendix for ranked $G^{(2)}$ of all provinces), where lighter colors indicate relatively smaller $G^{(2)}$. Hainan is excluded from the analysis of $G^{(2)}$, as it has no contiguous provinces. The mean value of $G^{(2)}$ for each province is 0.0243 with a standard deviation of 0.0120.

We compare the results of real data with that of the null model in order to understand the magnitude of $G^{(2)}$. In the null model \cite{gotelli2001research,berry2021leadership}, we randomly assign $G^{(1)}$ for each province and recalculate $G^{(2)}$. If the level of the gender pay gap in China is significantly influenced by the level of regional socio-economic development (and therefore the level of gender inequality in neighboring provinces is similar), the mean of $G^{(2)}$ of real data should be significantly smaller than that of the null model. Figure \ref{fig2}C compares the descending order of $G^{(2)}$ with the average values obtained from 1000 simulations of the null model (each simulation is arranged in descending order and the values of each rank are averaged in turn). Although $G^{(2)}$ of real data at the tail is slightly smaller than the null model, the overall difference is not significant. The mean of $G^{(2)}$ after 1000 simulations of the null model is 0.0253 with a standard deviation of 0.0018, and $G^{(2)}$ of real data is slightly smaller than that of the null model. However, in 1000 simulations, there are 227 times that the mean of $G^{(2)}$ of real data is larger than the null model, so the $p$-value is 0.227, which is not significant. We further conduct Mann-Whitney U test \cite{mann1947test} on the real and simulated values, and the $p$-value is 0.283, which is close to the direct estimation and also not significant. Thus the gender pay gaps do not show regional clustering.

We then test the correlation between the level of regional socio-economic development and the gender pay gap. Prior studies claim that economic and educational development could reduce gender inequality \cite{reskin2005including}. Specifically, economic development provides women with access to better-paying jobs in the non-agricultural sector \cite{duflo2012women}, including the tertiary industry in which women enjoy comparative advantages \cite{rosenzweig2013economic}. Educational opportunities for all could reduce gender differences in education and improve the competitiveness of women in the labor market \cite{hout2006we}. We use GDP per capita and the average number of college students per 100,000 persons as characteristic indices of the overall economic and educational level of a province, and analyze their relationships with the female-male ratio (see Figure S1 in SI Appendix). The results show that the economic development is barely correlated with the gender pay gap (Pearson’s $r$=-0.0389 and $p$-value$>$0.01 according to the Student's $t$-test), and the educational development is weakly correlated with the gender pay gap (Pearson’s $r$=-0.1245 and $p$-value$>$0.01 according to the Student's $t$-test). In addition, the data points are scattered, with many of them distant from the fitted regression line, so the reducing effect of economic and educational development on the gender pay gap should be small. Although higher educational development level is supposed to result in a smaller gender pay gap, we do not find support for it, as the educational inequality between men and women in China has already been significantly reduced with the compulsory education and China's emphasis on female education. There are ``more women than men” in higher education nowadays \cite{wei2013class}, but the gender pay gap is still big. Taken together, we believe that although development in economy and education can increase pay for both genders, it does not narrow the gender pay gap. In other words, by dominantly relying on economic and educational development, China seems to reach a plateau in shrinking the gender pay gap, as manifested in the widened gender pay gap in the last decade.

\begin{figure}[tbhp]
\centering
\includegraphics[width=0.7\linewidth]{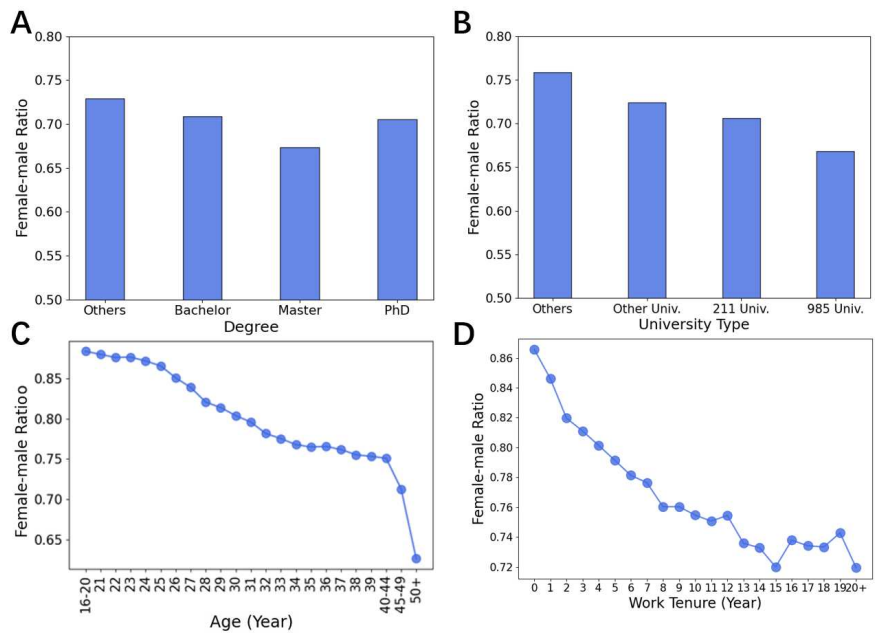}
\caption{The female-male ratios for job seekers with different (A) degrees, (B) university types, (C) age, and (D) work tenure.}
\label{fig3}
\end{figure}

At the micro level, we first examine the relationship between individual human resource endowments and the gender pay gap. Figure \ref{fig3} shows the female-male ratios of groups with different degrees, university types (elite or not), age, and work tenure. As shown in Figure \ref{fig3}A, the female-male ratio does not increase with the educational level, but it decreases as the university becomes more elite (see Figure \ref{fig3}B). Figure S2 (see SI Appendix) gives the absolute number of women's and men's pay with different educational levels. It is obvious that the pay of men with the same education level is still significantly higher than that of women. Women with higher degrees may averagely earn less than men with lower degrees. For example, the average pay of males holding a master degree is slightly higher than that of females with a PhD. Figure \ref{fig3}C and Figure \ref{fig3}D show the change of female-male ratio with age and work tenure (see also Figure S3 in SI Appendix for the change of pay with age and work tenure). We can see that the female-male ratio becomes significantly smaller with longer work tenure and among older employees. That is, the longer the work tenure, the larger the pay inequality. In sum, similar to the findings in the macro-level analysis, higher education and more work experience can increase individual earnings, but it widens the gender pay gap rather than reducing it.

Women with the same human capital earn less than men due to discrimination. We apply the Blinder-Oaxaca (BO) decomposition (see Materials and Methods for details) for resume data. The detailed results (see Figure S4A) show that human capital accounts for only 18.53\% of the gender pay gap, while the unexplained part, which can be potentially attributed to discrimination, accounts for 81.47\%. The results of the BO decomposition reconfirm our previous finding: the effect of improved female human capital on the gender pay gap reduction may be insignificant.

\begin{figure}[tbhp]
\centering
\includegraphics[width=.6\linewidth]{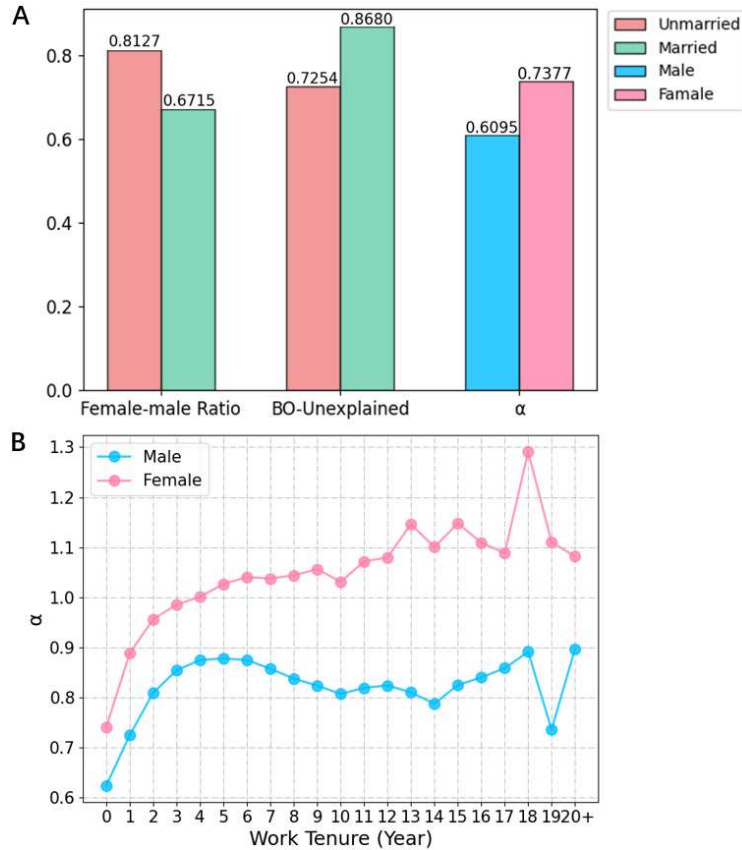}
\caption{(A) Comparisons between the female-male ratios for unmarried and married individuals, the unexplained part by BO decomposition for unmarried and married individuals, and $\alpha$ values for male and female. (B) The change of $\alpha$ as the increasing of work tenure for male and female.}
\label{fig4}
\end{figure}

Among factors not directly related to human capital, we pay particular attention to women’s marital status. The family constraint theory holds that the married women tend to undertake more family care responsibilities than men, which reduces their competitiveness and thus pay in the workplace \cite{becker1985human,bianchi2000anyone}. We divide all the individuals into two groups—the married and the unmarried, and analyze them separately. As shown in Figure \ref{fig4}A, the gender pay gap is significantly lower for the unmarried (the female-male ratio is 0.8127 for the unmarried and only 0.6715 for the married). The unexplained part, the part potentially attributed to discrimination, is significantly larger for the married (from 72.54\% to 86.80\%), indicating that married women suffer greater inequality and more discrimination than the unmarried ones. The detailed results of the BO decomposition for the unmarried and the married are shown in Figure S4B and S4C.

We define $\alpha$ as the ratio of the average pay of the married and unmarried individuals for male or female:

\begin{equation}
    \alpha = S_{unmarried}/S_{married},
\label{eqn3}
\end{equation}
where $S_{unmarried}$ is the average pay of unmarried individuals, and $S_{married}$ is the average pay of married individuals for male or female. As seen in Figure \ref{fig4}A, $\alpha$ is smaller than 1 for both male and female, namely the average pay of married individuals is higher than that of the unmarried (not only because those with higher pay are more likely to get married \cite{burstein2007economic,ravanera2007changes}, but also because married individuals tend to be older and have longer work tenure). However, married men enjoy more advantages over unmarried men ($\alpha$ for male is smaller than $\alpha$ for female). To remove the effects of age and work tenure, we group the data by work tenure. As seen in Figure \ref{fig4}B, the average pay of married men is significantly higher than that of unmarried men with different work tenure ($\alpha < 1$, consistent with the opinion that higher pay increases the likelihood of marriage), while the opposite is true for women who have work tenure over 5 years ($\alpha > 1$). The lower pay of unmarried women with a work tenure less than 5 years can likely be explained by the positive association between pay and marriage. Figure S5 (see SI Appendix) shows how $\alpha$ changes with age, which is similar to Figure \ref{fig4}B. The results above show that being married disadvantages working women —married women face more discrimination and earn less than men.

\section{Discussion}
Our research shows that China still faces great gender pay inequality, while at the same time, it hits a plateau to reduce it. Despite the Chinese government's continued efforts to improve its economy and education, it has not successfully shrinked the gender pay gap, although the average earnings for all have increased. The results of the BO decomposition further show that, there still exists a significant gender pay gap for women and men who have similar human capital, and a sizable part of the gap can be explained by women's marital status: married women have to face more discrimination in the labor market than unmarried women. According to the 2018 China Time Use Survey, unmarried men and women averagely spend 23 and 37 minutes per day on housework and family care, compared to 81 and 225 minutes for married men and women. As married women devote more time to their families, the time available for them for paid work is greatly restricted, thus affecting their labor returns \cite{south1994housework}. At the same time, due to the gender stereotype, women are perceived to be responsible for and good at housework, which further exacerbates the tension between women's dual roles as caregivers and breadwinners. This forces some women to choose lower paying occupations with limited advancement opportunities but more conducive to family care \cite{hook2010gender,cook2011harsh}. In short, family constraints have become an important cause of the gender pay gap in China. To reduce gender pay gap, the government should lighten the heavy family responsibilities on women's shoulders. We suggest the government to consider the following approaches: (i) introducing time policies, (ii) promoting care services, and (iii) providing financial support.

Time policies, represented by paid maternity and parental leave, are designed to balance women's time at home and at work. However, the generous and female-oriented time policies may lead to undesirable consequences, including raised costs of the companies and employers' hesitance to hire women. The policies would thus disencourage women to work and hinder their long-term career development \cite{amuedo2005motherhood}. Therefore, policy makers shoud target at both female and male employees and avoid to tie women to family care. Women should have equal opportunity as men, so that they can freely allocate their time to work or family. In China, the same or similar parental leave for both female and male employees can be piloted to foster a more equal gender division of family care responsibilities and reduce gender employment discrimination. Organizations are encouraged to replace some maternity and parental leave policies with longer flexible working time, combined with overtime pay incentives, to increase women’s willingness to return to work after childbirth.

Care services help to address family care responsibilities with paid professionals. In China, care work was once ``de-familized”, as publicly owned organizations took care of childcare and elderly care with the government bearing most of the costs. However, with the market-oriented reform, services such as childcare, elderly care, and housework have been commercialized. Family care has turn into a family responsibility or a commodity to be purchased. The lack of laws and regulations regarding the care service industry leaves many care workers with no vocational training and no deserved protection of their rights. As a result, high-quality services still lack. The limited supply also means they are luxury for the majority. Therefore, the government needs to boost the supply of all-inclusive care services, such as nurseries for children aged 0-3 years old and high-quality elderly care institutions. It is also necessary to set the standards and regulations of care services to ensure the service quality and protect care worker benefits. 

Financial support may serve as the most direct support. However, when it is provided on a household basis, families tend to maximize the benefits of the household as a whole, likely enhancing the employment willingness of the main earner. It may discourage the secondary earner in the household (usually the wife) to work, and solidify the existing gender division of family work \cite{azmat2010targeting}. Therefore, it is essential to ensure that women receive direct financial support if they are engaged in paid work. For example, individual tax policies can favor female employees, especially for working mothers. In addition, tax reduction for companies can be given based on the gender ratio, so that companies are encouraged to recruit more female employees.

The core target of care work, which is female-dominated, is to create, develop, and maintain human capabilities. It is a form of public goods characterized by the ``non-exclusivity” and ``non-competitiveness” at the supply and demand sides. As with all public goods, care work (especially childcare) can benefit the whole society, but cannot get deserved payment \cite{folbre1994children}. The market often fails when it comes to public goods. Placing the costs entirely on individuals or companies is not a good option \cite{england2020progress}. Especially in China, where people’s reproductive intention remains low and the gender pay gap is gradually widening, it is urgent for the government to make public investments and to introduce policies to raise fertility rates and encourage pay for care work. In doing so, it can possibly provide more equal employment opportunities for women without reducing their desire to become a mother. We believe that the analysis and conclusions of this research will help the Chinese government to formulate more targeted and precise policies. It will also be useful for countries and regions facing similar situations.

\section{Materials and Methods}
\subsection{Data Description}
The resume data of 10,318,484 job seekers is crawled from various online recruitment websites in 2015. The data includes basic personal information, educational experience, work experience, and expected position. Among them, gender, age, educational experience, marital status, last year salary, and work tenure (measured by the length of service) from work experience are utilized. The academic degree contains four levels: Ph.D., Master, Bachelor, and Others. The school where a job seeker earned the highest degree is categorized into four groups according to the Ministry of Education of China (\href{http://www.moe.gov.cn}{http://www.moe.gov.cn}) in 2015: (i) highest level ``985” universities (39 universities included in China's ``985 Program”, about 1.29\% of all universities in China), (ii) high level ``211” universities (Universities included in China's ``211 Program” are the second best graduate institutions next to ``985” universities. There are 115 universities excluding those also included in the ``985 Program”, about 3.81\% of all universities in China), (iii) other universities, and (iv) others. We label it as university type.

The resume data of a job seeker is used for further analysis if data on gender, age (over 16 years old), and last year salary are all available. This yields a data of 3,956,301 job seekers’ resumes, 2,373,962 of males and 1,582,339 of females. The basic statistics for resume data is shown in Table \ref{tab2}. In the analyses related to martial status and socio-economic development, we only consider resumes with sufficient demographic information, including gender, age, work tenure, degree, university type, last year salary, marital status, major, industry, position, the city of birth, the current city, and the city where the sought job is located, resulting in a smaller but still big data set composing of 753,616 job seekers' resumes. All the variables except last year salary and categorical variables are Z-standardized.

We retrieve data concerning province socioeconomics (e.g., GDP per capita and the average number of college students per 100,000 population) from ``China Statistical Yearbook (2016)”, which summarizes the statistics of China’s socio-economic status in 2015.

\subsection{Blinder-Oaxaca Decomposition}
The Blinder-Oaxaca (BO) decomposition \cite{blinder1973wage,oaxaca1973male} is one of the most classic methods of pay gap decomposition, which decomposes pay gap between gender groups into the explained part caused by individual differences, and the unexplained part potentially caused by gender discrimination. Regression of the logarithm of salary yields the following formula:

\begin{equation}
\begin{aligned}
    &\log(S_M) = \beta_M X_M+\epsilon_M, \\
    &\log(S_F) = \beta_F X_F+\epsilon_F,
\end{aligned}
\end{equation}
where $S$ represents the average annual salary, $X$ is the influencing factors of salary (i.e. the characteristic matrix), $M$ and $F$ are short for male and female. Then, the $GPG$ ($GPG=(S_M-S_F)/S_M$) can be expressed as:

\begin{equation}
    GPG \approx (\bar{X}_M-\bar{X}_F)\beta_M+(\bar{\beta}_M-\bar{\beta}_F)\bar{X}_F,
\end{equation}
where $(\bar{X}_M-\bar{X}_F)\beta_M$ (explained part) represents the gap due to individual differences (such as educational level, work tenure, etc.) assuming there is no gender discrimination; and $(\bar{\beta}_M-\bar{\beta}_F)\bar{X}_F$ (unexplained part) is the gap potentially resulted from gender discrimination.

\section*{Acknowledgments}
We thank Jian Gao and Zhong-Tao Yue for providing the raw data. This work was supported by the National Natural Science Foundation of China (Grant No. 11975071), and the Ministry of Education of Humanities and Social Science Project (Grant No. 21JZD055).


\end{document}